\documentclass[a4paper,11pt]{article}
\pdfoutput=1
\usepackage{jheppub} 
\usepackage{graphicx}
\usepackage{amssymb,amsmath,mathrsfs}
\usepackage{cancel}
\usepackage{subfig}
\usepackage{hyperref}
\usepackage{verbatim}

\def\bea{\begin{eqnarray}}
\def\eea{\end{eqnarray}}
\def\beq{\begin{equation}}
\def\eeq{\end{equation}}

\newcommand{\bear}{\begin{array}}
\newcommand{\ear}{\end{array}}

\def\OMIT#1{{}}
\newcommand{\lsim}{\mathrel{\rlap{\lower4pt\hbox{\hskip1pt$\sim$}}
    \raise1pt\hbox{$<$}}}         
\newcommand{\gsim}{\mathrel{\rlap{\lower4pt\hbox{\hskip1pt$\sim$}}
    \raise1pt\hbox{$>$}}}         

\newcommand{\be}{\begin{eqnarray}}
\newcommand{\ee}{\end{eqnarray}}
\newcommand{\ba}{\begin{eqnarray}}
\newcommand{\ea}{\end{eqnarray}}

\newcommand{\Ncp}{N_{c^\prime}}
\newcommand{\bbZ}{\mathbb{Z}}
\newcommand{\lameff}{\lambda_{\mathrm{eff}}}
\newcommand{\sthr}{s_{\mathrm{thr}}}
\newcommand{\Mpl}{M_\mathrm{Pl}}
\newcommand{\TeV}{\mathrm{TeV}}
\newcommand{\GeV}{\mathrm{GeV}}

\newcommand{\MSbar}{\overline{\mathrm{MS}}}
\newcommand{\LamQCD}{\Lambda_{\mathrm{QCD}}}
\newcommand{\LamQCDp}{\Lambda_{\mathrm{QCD}^\prime}}
\newcommand{\mrhop}{m_{\rho^\prime}}
\newcommand{\Mbi}{M_{\mathrm{bi}}}

\begin{document}
\title{Solving the Wrong Hierarchy Problem}
\author[a]{Nikita Blinov,}
\author[b]{Anson Hook}
\affiliation[a]{SLAC National Accelerator Laboratory, 2575 Sand Hill Road, Menlo Park, CA, 94025, USA}
\affiliation[b]{Stanford Institute for Theoretical Physics, Stanford University, Stanford, CA 94305, USA}
\emailAdd{nblinov@slac.stanford.edu}
\emailAdd{hook@stanford.edu}
\abstract{
Many theories require augmenting the Standard Model with additional scalar
fields with large order one couplings. We present a new solution to the
hierarchy problem for these scalar fields. We explore parity- and
$\mathbb{Z}_2$-symmetric theories where the Standard Model Higgs potential has
two vacua.  The parity or
$\mathbb{Z}_2$ copy of the Higgs lives in the minimum far from the origin while
our Higgs occupies the minimum near the origin of the potential. 
This approach results in a theory with multiple light scalar fields but with
only a single hierarchy problem, since the bare mass is tied to the Higgs mass
by a discrete symmetry.
The new scalar does not have a new hierarchy problem  associated with it
because its expectation value and mass are generated by dimensional
transmutation of the scalar quartic coupling. 
The location of the second Higgs minimum is not a free
parameter, but is rather a function of the matter content of the theory. As a
result, these theories are extremely predictive. We develop this idea in 
the context of a solution to the strong CP problem. We show 
this mechanism postdicts the top Yukawa to be within $1 \sigma$ of 
the currently measured value and predicts scalar color octets with masses in the range 9-200 TeV.}

\date{\today}
\preprint{SLAC-PUB-16518}
\maketitle

\setlength{\parindent}{4ex}

\section{Introduction}
\label{sec:intro}

The Standard Model (SM) Higgs boson mass is sensitive to ultraviolet (UV) physics.
This is the hierarchy problem.  Traditional solutions to the hierarchy problem
have postulated the existence of new particles that appear at low scales to
cancel the quadratic dependence on UV physics.  Recently there has been renewed
focus on non-traditional solutions to the hierarchy problem, such as 
the multiverse, the relaxion~\cite{Graham:2015cka}, and
$N$Naturalness~\cite{God_knows_when}. In these
solutions, the Higgs mass appears to have been fine
tuned to be small. 

Aside from the hierarchy problem, the SM has other issues that require
explanations, such as the origin of flavor, baryogenesis and the strong CP
problem. Solutions to these problems often introduce new heavy scalar degrees
of freedom. These scalar fields also suffer from quadratic sensitivity to UV
physics.  Usually these ``wrong'' hierarchy problems are resolved in the same way as
the electroweak hierarchy. In this work, we present a new solution to this
secondary hierarchy problem that operates when the naturalness of the
electroweak scale is explained using one of the non-traditional mechanisms
mentioned above.

Our approach employs dimensional transmutation through 
the Coleman-Weinberg mechanism to dynamically generate a
new hierarchy of scales~\cite{Coleman:1973jx}. 
Standard use of dimensional transmutation to 
stabilize scalar mass relies on gauge
couplings becoming strong and gives rise to the large separation between
the QCD and the electroweak scales. When applied to the usual naturalness
problem, it leads to technicolor and its variants. There is another kind of
dimensional transmutation which occurs in the SM and involves the Higgs scalar quartic
instead of a gauge coupling. This effect is well known through the
renormalization group evolution (RGE) of the Higgs self-coupling $\lambda$,
which generates a maximum in the Higgs potential and an instability at large
field values~\cite{Degrassi:2012ry,Buttazzo:2013uya,Bednyakov:2015sca}. The
dimensional transmutation aspect of this mechanism was emphasized in
Ref.~\cite{Einhorn:2007rv}. We utilize this effect to solve the secondary
hierarchy problem discussed above. A similar approach was used to 
dynamically generate the Planck scale in Ref.~\cite{Salvio:2014soa}.

Our Higgs field occupies the very long lived electroweak minimum, so the
apparent run-away of the potential at large field values $h$ is irrelevant.
However, in parity- or $\mathbb{Z}_2$-symmetric theories, there is a second
Higgs with a similar potential. Additional matter can radiatively stabilize
this potential at large $h$ and generate a second minimum that is much deeper
than the electroweak one. If the second Higgs occupies this true vacuum, it can
provide a new heavy scale that can be used to solve any of the aforementioned
problems. This new scale is generated by dimensional transmutation and so there
is no new hierarchy problem.

In order for the dimensional transmutation of the scalar quartic to be important, 
the relevant mass term must be small. We use a discrete 
symmetry to link this new scalar mass with the Higgs mass. In this way, the
mechanism that solves the standard hierarchy problem also ensures the mass
parameter of the new scalar field remains small. 
Thus our starting point is
a $\mathbb{Z}_2$ or parity  symmetric theory, 
where the symmetry operation exchanges SM fields for their ``mirror'' partners.\footnote{Similar $\bbZ_2$-symmetric
constructions have been used, for example, to explain dark matter and the
baryon asymmetry of the universe, naturalness of the weak scale and to solve
the strong CP problem -- see
Refs.~\cite{Okun:2006eb,Chacko:2005pe,An:2009vq,Berezhiani:2000gh,Hook:2014cda,D'Agnolo:2015uta,Foot:2014mia} and
references therein.}
The discrete symmetry guarantees that the renormalization group running of
couplings in the mirror sector is identical to the SM at high scales. This ensures that
the potential for the $\bbZ_2$ partner of the Higgs $H^\prime$ 
can be determined by studying the
potential of the SM Higgs boson $H$. 

The tree-level $\bbZ_2$ symmetric scalar potential has several minima.  
These minima have many problems.  There are new light particles and the new scalar 
is at the weak scale or lower, making it too light for the solution of, e.g., 
flavor or strong CP problems.  
Thus it is impossible to get a 
hierarchy between $h=\sqrt{2}\langle H \rangle$ and 
$h^\prime = \sqrt{2}\langle H^\prime \rangle$ at \emph{tree-level}.
Renormalization group evolution is necessary for $h^\prime$ to be separated from the weak scale.
Because logarithmic RGE requires a large amount of running to be effective, this new
scale will be parametrically larger than the weak scale.

There is a simple way to obtain the large amount of running needed to generate
a new scale.  If the masses of the scalars are parametrically below the cutoff
$\Lambda$, then loop corrections to the 
effective potential may create a new minimum where $\Lambda \gg h^\prime \gg h$.
In order for the mass term $m_H$ to not be important in determining the location of the minimum, 
we necessarily have $h' \gg m_H$.\footnote{Another way for RG running of the
  Higgs quartic to play a role is in tuned Twin Higgs
  models~\cite{Chacko:2005pe}.  As it is similar to what we consider here, we leave this example as an exercise for the reader.}

Because all of the SM parameters have been measured, the location of the second
vacuum is determined by the matter content of the theory. Thus the new heavy scale
is completely determined without the introduction of new dimensionful
parameters, leading to a very predictive framework. The radiatively generated
scale is most sensitive to the strong coupling $\alpha_s$ and top quark Yukawa
$y_t$. Precise measurements of these quantities will be necessary to accurately
determine the new energy scale.

In this work we illustrate this mechanism by considering an $\eta'$ solution to
the strong CP problem~\cite{Hook:2014cda}. A $\mathbb{Z}_2$ symmetry and new
massless quarks allow the QCD $\theta$ angle to be rotated away. Dimensional
transmutation naturally determines the vacuum expectation value (vev) of the $\bbZ_2$ Higgs copy 
to be below the Planck scale, fixes the masses of colored states 
(originating from the new quarks) 
and precisely predicts the top quark Yukawa to be within $1\sigma$
of the currently measured value.

This paper is organized as follows. 
In Sec.~\ref{sec:cpsol}, we present a solution to the wrong hierarchy
problem in the context of an $\eta'$ explanation of the strong CP problem finding interesting
experimental predictions.
In Sec.~\ref{sec:dimtrans} we discuss the details of the Higgs potential in this solution
to the strong CP problem.
We briefly consider further applications of this 
mechanism and conclude in Sec.~\ref{sec:conclusion}. Technical details
are collected in the Appendices.

\section{Hierarchies and the Strong CP Problem}

In this section, we illustrate our solution to the wrong hierarchy problem in
the context of a predictive $\eta'$ explanation of the strong CP
problem.  Discrete symmetry-based solutions of strong CP involve a new scale where
this symmetry is spontaneously broken.  Stabilizing this new scale
with traditional solutions such as supersymmetry (SUSY) is 
difficult due to the plethora of new
phases introduced by SUSY breaking and due to unsuppressed renormalization group running
of $\theta$~\cite{Albaid:2015axa,Dine:2015jga}. 
In contrast, our solution does not introduce new CP violating phases
and preserves the small running of $\theta$ present in the SM.

The minimal $\eta'$ solution to the strong CP problem involves two copies of
the SM related by a $\mathbb{Z}_2$ symmetry. 
In addition, there are new massless vector-like fermions, $\psi$ and
$\psi^c$, which are bifundamentals under our QCD and QCD$'$ (the $\mathbb{Z}_2$
copy of QCD).\footnote{Mirror sector quantities are denoted 
with primes.}  In the absence of a vector-like mass, an anomalous $U(1)$ rotation 
of $\psi$ and/or $\psi^c$ is used to set the sum of the two theta angles to zero, 
while their difference is zero by the $\mathbb{Z}_2$ symmetry. 
Thus the strong CP problem is solved in the unbroken $\bbZ_2$ limit.

Non-observation of new light states suggests that $\mathbb{Z}_2$ breaking must occur and that the 
mirror sector's Higgs vev must be large, potentially introducing a new hierarchy problem. 
In our solution, the $\mathbb{Z}_2$
symmetry ensures that the tree-level mirror Higgs mass is small. 
As mentioned in the introduction, the $\mathbb{Z}_2$ breaking
then comes from the RG running of the Higgs quartic so that the mirror Higgs
obtains a large physical mass and vev without introducing a new hierarchy problem.

\subsection{A Predictive $\eta'$ Solution to the Strong CP Problem}
\label{sec:cpsol}

\begin{figure}[t]
\begin{center}
\includegraphics[width=0.45\textwidth]{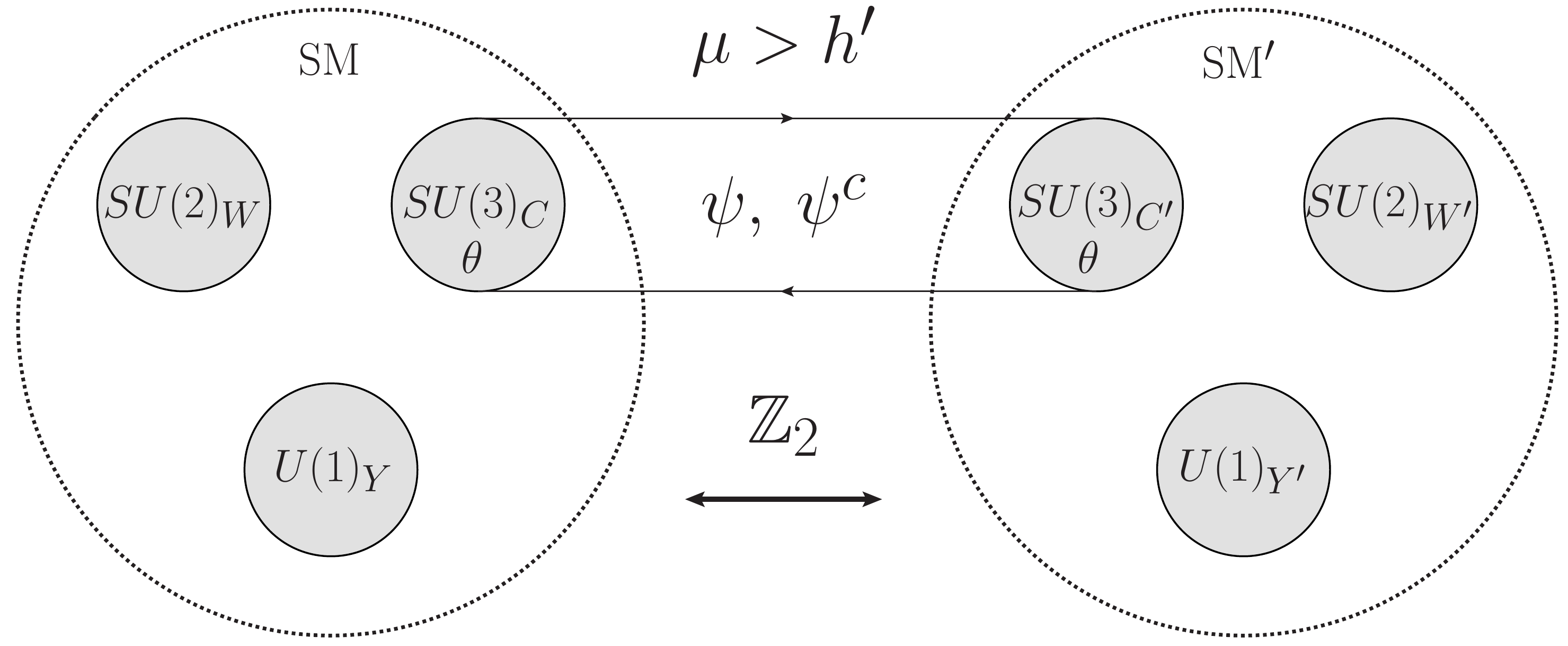}
\includegraphics[width=0.45\textwidth]{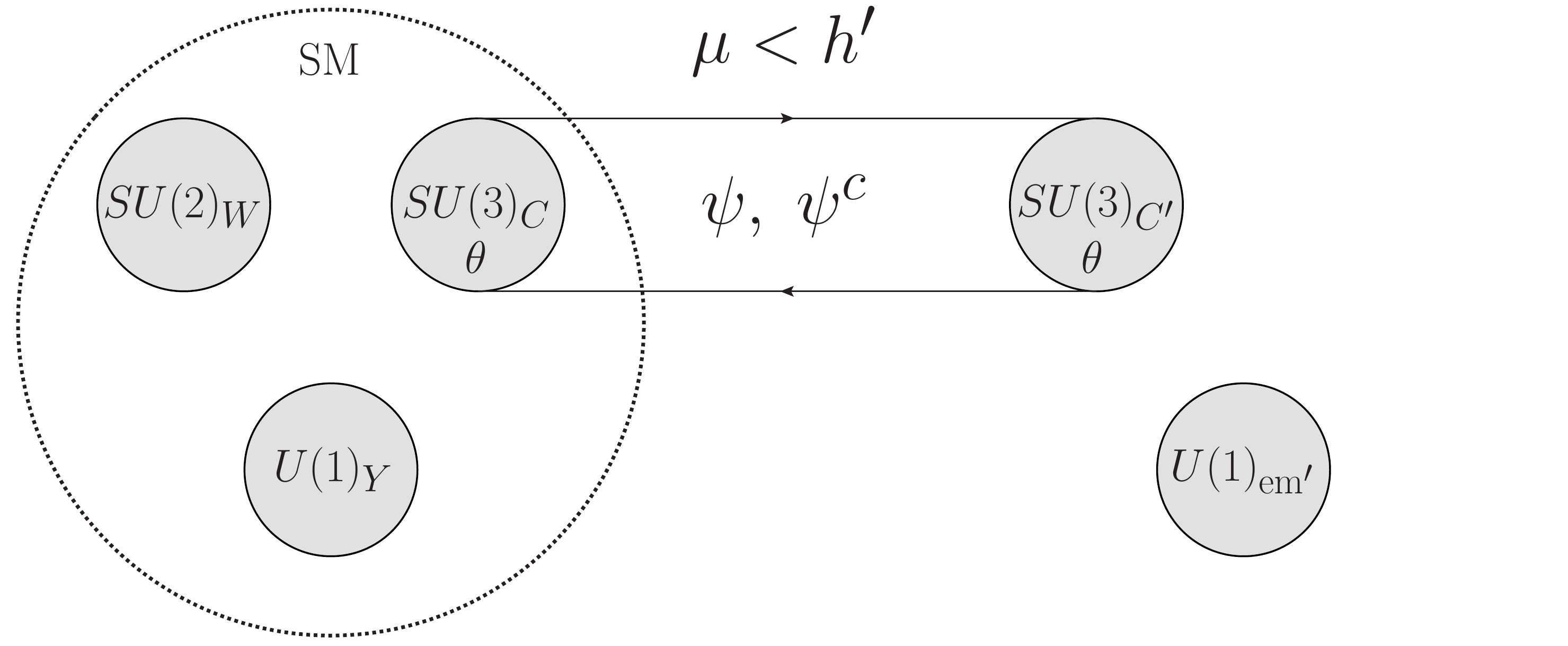}
\caption{A pictorial representation of the $\eta'$ solution to the strong 
CP problem at various RG scales $\mu$. The left (right) diagram shows the 
model above (below) the $\bbZ_2$ breaking scale $h'$.} \label{Fig: model}
\end{center}
\end{figure}

As will be shown in the next section, 
the potential for each Higgs naturally contains two vacua in this scenario. 
If both Higgs bosons are in the same minimum, then the strong CP problem is solved
by a chiral rotation of a massless quark. This solution is
observationally excluded due to the presence of three new massless vector-like
quarks.  However, if the two Higgs bosons occupy different vacua, then the
particles in the mirror sector can be much heavier than any SM states.  Due to the 
larger $h^\prime$ vev, the quarks and leptons in the mirror sector are
integrated out at much larger scales.  After all of the new massive particles
have been integrated out, the matter content is simply the
SM augmented by the gauge group $SU(3)_{C'}$ and the associated bifundamentals $\psi$
and $\psi^c$. At this point, because the $\mathbb{Z}_2$ symmetry has been
broken by $h\neq h^\prime$, the effective QCD angle 
$\overline \theta$ is reintroduced by renormalization group effects. 
However, as shown in Refs.~\cite{Ellis:1978hq,Khriplovich:1993pf}, 
the RG effects in these types of models are
negligible. This model is shown schematically in Fig.~\ref{Fig: model} for 
RG scales $\mu \gg h^\prime$ (left) and $\mu \ll h^\prime$ (right).

After QCD$'$ confines, there are no more massless degrees of freedom in the
infrared (IR) and the theory is simply the SM with extra massive colored states, 
the pseudo-Goldstone bosons of QCD$'$. As emphasized in
Ref.~\cite{Hook:2014cda}, if the $\theta$ angle has not been rotated to
zero, dynamic relaxation of the the $\eta''$ (the $\eta'$ boson of QCD$'$) sets our
$\overline \theta$ to zero.  More explicitly, the IR effective theory of the
$\eta''$ after QCD$'$ confines is 
\bea
\label{Eq: chiral}
\mathscr{L} = \frac{g^2}{32 \pi^2} \left ( \overline \theta - \frac{\eta''}{f_{\eta''}} \right ) G \tilde G + \frac{m_{\eta''}^2}{2} \left ( \eta'' - f_{\eta''} \overline \theta' \right )^2 + \cdots
\eea
where the ellipsis stands for other terms in the chiral Lagrangian. 
Integrating out the $\eta''$ boson and using the $\mathbb{Z}_2$
symmetry ($\overline \theta = \overline \theta'$), we find that our theta angle
has been set to zero.

The presence of $\psi$ and $\psi^c$, three new flavors of $SU(3)_C$
quarks, introduces a small stabilizing effect on the RG running of the Higgs
quartic and generates a second minimum in the Higgs potential for certain ranges of
SM parameters.  The location of this true minimum determines the confinement 
scale of QCD$^\prime$ and therefore the mass scale of states associated with $\psi$ and
$\psi^c$.  However, the scale where $\psi$ and $\psi^c$ are integrated in 
determines the true minimum.  Therefore finding the true minimum is a
recursive process that leads to somewhat counter-intuitive results.  We 
leave a detailed discussion of the Higgs potential and 
the determination of the bifundamental mass scale to 
Sec.~\ref{sec:dimtrans} and simply summarize the results here.

\begin{figure}[t]
  \centering
  \includegraphics[width=0.48\textwidth]{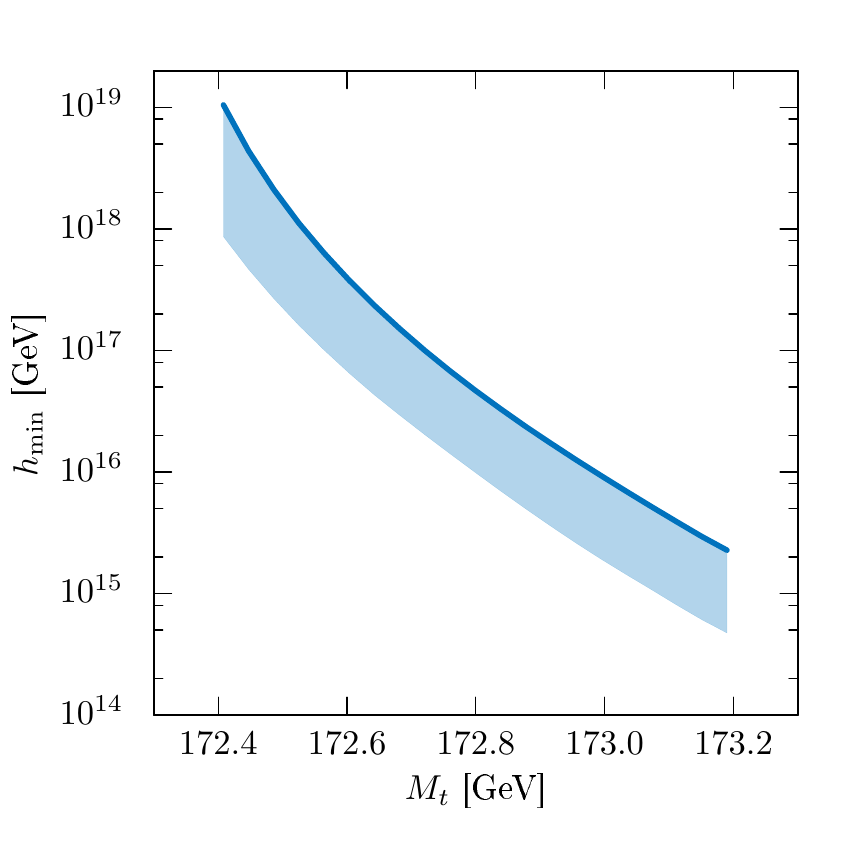}
  \includegraphics[width=0.48\textwidth]{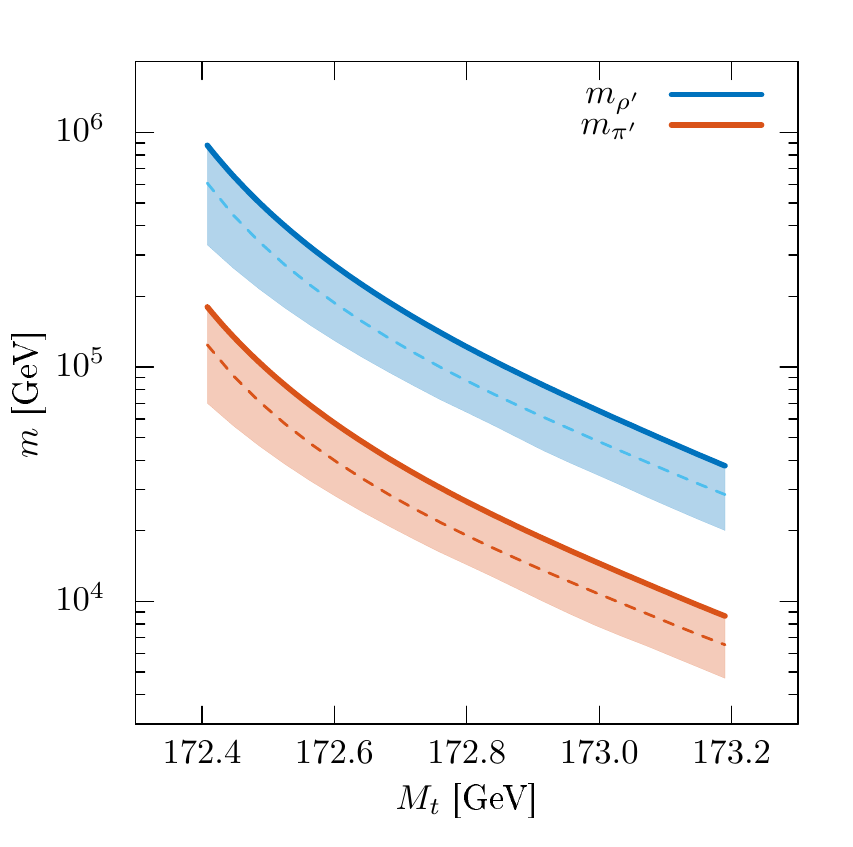}
  \caption{(Left) Dependence of the mirror Higgs VEV on the top quark mass $M_t$.
  (Right)  The mass of the color octet $\rho'$ (upper line) and $\pi'$ mesons
(lower line) as a function of $M_t$.  The uncertainty bands shown are
generated by varying the matching scale associated with running the QCD
coupling through the strongly coupled regime of QCD$^\prime$. The solid (dashed) line 
in each band corresponds to using the four (one) loop
QCD$^\prime$ beta function to determine the confinement scale $\LamQCDp$. 
Below the lower range of $M_t$ shown the mirror Higgs vev 
moves above the Planck scale, while above the highest $M_t$ shown 
no self-consistent solution for bifundamental mass scale exists.
}  \label{Fig: hmin} 
\end{figure}

We find that the second minimum only exists for a
small range of parameters. As the Higgs potential is most sensitive to the top
quark Yukawa, we hold other parameters at their central measured values and vary the
top quark mass $M_t$. The corresponding $\MSbar$ quantities also vary with $M_t$ 
due to loop corrections~\cite{Degrassi:2012ry,Buttazzo:2013uya,Bednyakov:2015sca}. 
The mirror Higgs vev as a function of $M_t$ is shown
in the left plot of Fig.~\ref{Fig: hmin}. 
The determination of the uncertainty band is discussed in 
Sec.~\ref{sec:bifunmass} and Appendix~\ref{sec:strong_threshold}.
The top quark mass is postdicted to be in the range
\bea
172.4 \, \text{GeV} < M_t < 173.2 \, \GeV.
\label{eq:mt_range}
\eea
This is to be compared with the measured value~\cite{Agashe:2014kda} 
\bea
M_t = 173.34 \pm 0.87 \, \GeV.
\eea
The solution to the strong CP problem has accurately postdicted the top quark mass
to within $1\sigma$! Note that this postdiction depends on the representation of the
$\psi^{(c)}$ states. Thus, precise measurements of $M_t$ 
(such as those possible with the ILC~\cite{Fujii:2015jha}) can probe 
the UV content of this theory. 

The small range of allowed top quark masses 
can be understood as follows.
First, note that the smaller the radiatively-generated vev is, the earlier $\psi$ and $\psi^c$ 
are integrated into the RGEs and the larger their effect on the potential at high 
field values is. Therefore to counteract a larger $y_t$, $\psi$ and $\psi^c$
must appear sooner and thus the vev must decrease. This explains the 
negative slopes in the vev and the QCD$^\prime$ confinement scale as a function
of $M_t$ shown in Fig.~\ref{Fig: hmin}. Above the upper bound in 
Eq.~\ref{eq:mt_range}, the bifundamentals are so light 
that their indirect effect is enough to 
completely stabilize the electroweak vacuum: the second minimum disappears. 
Below the lower bound in
Eq.~\ref{eq:mt_range}, the second Higgs minimum is pushed above the Planck
scale.  
This can be understood by noting that for lower $y_t$ the SM potential is
nearly stable, so $\psi^{(c)}$ must be integrated in at a high scale. 
As $y_t$ is lowered and the scale of $\psi^{(c)}$ is increased, 
the stabilization of the potential occurs at ever higher scales, eventually
passing $\Mpl$.  These arguments are made precise in the following section.

The lightest new particles in the theory come from the confinement of QCD$'$.
Confinement is accompanied by the spontaneous breaking
of global symmetry $SU(3)_L \times SU(3)_R \rightarrow SU(3)_D$, 
associated with the massless bifundamentals $\psi$ and $\psi^c$ (recall 
that \emph{all} of the mirror quarks have been integrated out).\footnote{The chiral symmetry breaking pattern depends on the representation of
  $\psi^{(c)}$. We work with bifundamentals of $SU(3)_C\times SU(3)_{C^\prime}$
to make use of existing QCD results.}
Analogously to QCD, the spectrum consists of an octet pseudo-Goldstone
bosons $\pi'$ and other mesons.
The chiral symmetry is also explicitly
broken by the gauging of QCD $SU(3)_C\subset SU(3)_L \times SU(3)_R$, so $\pi^\prime$ is 
a color octet.
Thus, much like the charged pions, the 
$\pi'$ obtain a mass from gluon loops. 
The confinement scale of QCD$'$ is determined by $h'$.  Because $h' >
h$, the mirror quarks are integrated out sooner and the confinement scale is
larger than in QCD.  The mass of the scalar color octet as a function of the top
quark mass is shown in the right plot of Fig.~\ref{Fig: hmin}.  We find that these new particles
are predicted to have a mass
\bea
9 \, \TeV < m_{\pi'} < 200 \, \TeV,
\eea
where the range is set by the precise value of $M_t$.
This new scalar color octet decays through the Wess-Zumino term 
into a pair of gluons, much like how the $\pi^0$ decays into a pair of photons.  
Single $\pi^\prime$ production can occur through the same interaction, 
leading to a resonant dijet signature at a hadron collider.
While these states are too heavy to be produced at the 
LHC~\cite{Han:2010rf,Chen:2014haa,Cacciapaglia:2015eqa}, 
a future 100 TeV collider will be able to probe a 
significant portion of this mass range~\cite{Arkani-Hamed:2015vfh}.
We reserve the detailed study of collider signatures of 
this scenario to a future work.

Solutions to the strong CP problem are 
sensitive to UV physics through higher dimensional operators. 
These problems for axions~\cite{Kamionkowski:1992mf} and 
the Nelson-Barr mechanism~\cite{Dine:2015jga} are well known.
In the present case, once $h\neq h^\prime$, there is no symmetry
protecting the QCD angle and it can again become non-zero.  In our model
the RG effects are negligible and $\overline \theta$ does not run very much.
The leading contribution to $\overline \theta$ comes from higher dimensional
operators such as
\bea
\delta \mathscr{L} \supset Y_u H Q u^c \frac{H H^\dagger}{\Mpl^2} + Y_u H' Q' u'^c \frac{H' H'^\dagger}{\Mpl^2}. 
\label{eq:higher_dim}
\eea
Using a chiral rotation, we see that this changes $\theta$ ($\theta'$) by $\sim
H H^\dagger / \Mpl^2$ ($H' H'^\dagger / \Mpl^2$).  When the two Higgs vevs become
unequal, these operators lead to a non zero theta angle at low energies
\bea
\overline \theta \sim \frac{h'^2 - h^2}{\Mpl^2} \gtrsim 10^{-8}
\eea
for the range of $h'$ vevs shown in Fig.~\ref{Fig: hmin}. For 
$\mathcal{O}(1)$ Wilson coefficients for operators in Eq.~\ref{eq:higher_dim}, 
this is two orders of magnitude bigger than the current best limits from 
neutron~\cite{Baker:2006ts} and mercury electric 
dipole moment searches~\cite{Griffith:2009zz,Engel:2013lsa}. 
The remaining tuning of $\overline \theta$ can be alleviated further with larger 
representations (or more families) of the connector states $\psi^{(c)}$, 
which would lower the mirror Higgs vev $h'$.

\subsection{Cosmology with a Mirror Sector}

We conclude this section with a brief description of bounds on the
reheat temperature of the universe.  Because of the mirror baryon and electron
number, the lightest stable particles of the other sector 
(their electrons, up quarks, and Dirac neutrinos) should
never be in thermal equilibrium or they would overclose the
universe.\footnote{This can be easily seen by the fact that they are
parametrically heavier than the weak scale where the correct thermal relic
density would be obtained.} Even in the most constraining situation, this
only limits the reheat temperature to be beneath $10^{10}$ GeV, 
corresponding to the mass of the lightest charged mirror states. The glueballs
and $\pi'$ of QCD$^\prime$ decay to SM hadrons before nucleosynthesis. A stronger 
constraint on the reheat temperature can come from overclosure due to 
stable $\psi$-baryons if their annihilation rate is too small. Alternatively, 
these states can play the role of dark matter if $\LamQCDp$ is low enough 
or if annihilation is otherwise enhanced, e.g., through a resonance.

The $\bbZ_2$ neutrinos are either extremely heavy (Majorana masses),
$\sim 100$ GeV in mass (Dirac masses), or can even act as the right handed
neutrinos through the higher dimensional operator~\cite{Akhmedov:1992hh} 
\bea
\mathscr{L} \supset \frac{H L H' L'}{\Mpl}.
\eea
Given bounds from over production of mirror electrons and quarks, the mirror
neutrinos are never in thermal equilibrium.  Dirac mirror neutrinos could be
dark matter depending on how they are produced.

Even in the absence of mirror quarks and leptons, we cannot reheat above the
scale $h' \gtrsim 10^{15}$ GeV as at that point thermal effects would stabilize
the scalar field origin for both sectors.  Both Higgses would live at the origin and the
$\mathbb{Z}_2$ symmetry would be restored resulting in a $\mathbb{Z}_2$
symmetric universe.

\section{Dimensional Transmutation from the Higgs quartic}
\label{sec:dimtrans}

In this section we discuss dimensional transmutation using the Higgs quartic.  A second
minimum in the Higgs potential is already generated in the SM for central
values of the Higgs boson and top quark masses and the strong coupling
$\alpha_s$. This is the observation that the electroweak vacuum is metastable with
a very long lifetime~\cite{Degrassi:2012ry,Buttazzo:2013uya,Bednyakov:2015sca}.
The expectation value of the Higgs field in the new minimum is 
far above the Planck scale where the SM is not valid~\cite{Andreassen:2014gha}. 
However, in the
presence of additional matter this minimum can move to sub-Planckian field
values or disappear completely. The latter possibility has been considered 
by many authors and is arranged by adding new matter that 
alters the RG evolution of $\lambda$, see, e.g., Refs.~\cite{Chao:2012mx,Chen:2012faa,EliasMiro:2012ay,Lebedev:2012zw,Espinosa:2015kwx}. 
In contrast, we will require that the SM vacuum is only metastable 
with the true minimum at sub-Planckian field values which can be 
studied using ordinary field theoretic methods.  As we are considering the case where
the Higgs is even more stable than in the SM, it is clear that the  lifetime of the metastable vacua
will be longer than the age of the universe.

\subsection{Tree-Level Potential in the $\bbZ_2$-symmetric Standard Model}

Our starting point is
a $\mathbb{Z}_2$ or parity  symmetric theory, as discussed in Sec.~\ref{sec:intro}. 
The parity operation exchanges SM fields and gauge 
group for their $\bbZ_2$ partners, denoted by primed quantities.
This symmetry ensures that the behavior of the effective potentials 
for the two Higgs states is the same at large field values.
The tree-level $\bbZ_2$ symmetric scalar potential is
\beq
V_0 = -m_H^2 (|H|^2 + |H^\prime|^2) + \lambda (|H|^4 + |H^\prime|^4) + \delta |H|^2 |H^\prime|^2,
\eeq
with $H^{(\prime)}$ having the vacuum expectation value $h^{(\prime)}/\sqrt{2}$. We will consider 
the case where loop corrections to this potential are important.  
As a result, the Higgs obtains a second
minimum at scales far above the weak scale but below the Planck scale. 
For simplicity we also set $\delta \ll g_1, g_2, y_t$ at a high scale so that its impact on the 
RG running of the quartic is negligible; this coupling will
change due to RG evolution in the presence of matter charged under both 
copies of the gauge group, but at a very high loop order.  This restriction 
allows us to consider the effective potentials of SM and SM$^\prime$ separately in the 
following sections. When $H'$ obtains an 
expectation value the tree-level Higgs mass is shifted by $\delta h^{\prime 2}/2$, which can be
large even if $\delta$ is small. 
This dynamical contribution to $m_H^2$ is subleading with respect to the 
usual naturalness problem and is dealt with by the ``non-standard'' solutions 
to the electroweak hierarchy discussed in the introduction.

\subsection{Renormalization Group and the Effective Potential}

Dimensional transmutation with the Higgs quartic $\lambda$
happens at large field values. In order to reliably study the potential in 
this regime we use the renormalization group-improved 
potential, where the couplings are evaluated at the scale of the field value. 
We work in the two-loop approximation both for the potential and the RG 
evolution of the coupling constants. This amounts to a 
next-to-leading log resummation~\cite{Kastening:1991gv,Bando:1992np}.\footnote{Next-to-leading 
  log accuracy is obtained already with the one loop potential improved by two loop 
beta functions.} 

At high field values where the mass parameter can be neglected, 
the SM Higgs potential can be cast into the tree level-like 
expression~\cite{Casas:1994qy}
\beq
\label{Eq: eff pot}
V(h)\approx \frac{\lambda_{\mathrm{eff}}(h)}{4}h^4,
\eeq
where the effective quartic interaction $\lameff$ encodes the loop corrections to the potential
\beq
\lameff(h) = \lambda(\mu=h) + \frac{4}{h^4} \left(V_1(\mu=h) + V_2(\mu=h)+\dots\right),
\eeq
and $V_1(\mu)$ and $V_2(\mu)$ are the one- and two-loop contributions 
to the effective potential.
All couplings in Eq.~\ref{Eq: eff pot} are evaluated at the scale $\mu=h$.
In the one- and two-loop potentials above we include the contributions of 
$W$ and $Z$ bosons, the top quark, the Higgs and its Goldstone modes. 
The two-loop potential has been evaluated in the SM~\cite{Ford:1992pn} 
and in general theories~\cite{Martin:2001vx} in the Landau gauge. We validated our implementation 
against the publicly available code \texttt{SMH}~\cite{Martin:2014cxa}. 
The evaluation of the general two-loop beta functions and anomalous dimensions 
of Refs.~\cite{Machacek:1983tz,Machacek:1983fi,Machacek:1984zw,Luo:2002ti} has 
been automated in Refs.~\cite{Staub:2013tta,Lyonnet:2013dna}. Explicit 
expressions for the SM potential and beta functions can found in, 
e.g., Ref.~\cite{Buttazzo:2013uya}.

\begin{figure}
  \centering
  \includegraphics{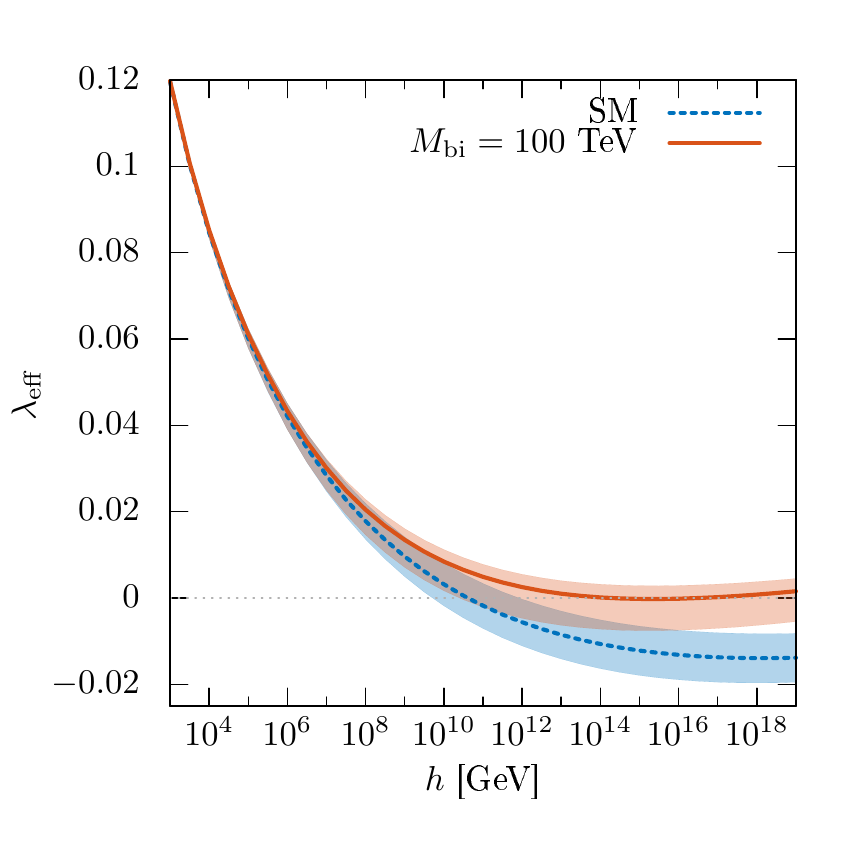}
  \caption{The effective quartic coupling as a function of the Higgs VEV for the SM field content and for a theory with additional $N_{c^\prime}=3$ new vector-like states charged under QCD that are integrated in at 100 TeV. The bands around each line correspond to varying $M_t$ within $1\sigma$. 
  \label{fig:lameff}}
\end{figure}

The potential is renormalized in $\MSbar$ at $\mu=M_t$, where $M_t$ is the top 
quark pole mass. The corresponding $\MSbar$ SM couplings at this scale are given in Ref.~\cite{Buttazzo:2013uya}
in terms of observables, including $M_t$ itself. In Fig.~\ref{fig:lameff} we 
show the effective quartic coupling as a function of $h$ 
for central values of SM parameters as the dashed line. The blue band 
corresponds to varying $M_t$ within its $1\sigma$ uncertainty. Note
that at large field values $\lameff<0$ corresponding to a potential 
that is unbounded from below up to the Planck scale.

As discussed in Sec.~\ref{sec:cpsol}, our example solution to the strong CP 
problem requires the addition of
massless bifundamental quarks transforming as $(\mathbf{3},\overline{\mathbf{3}})$ of $SU(3)_C\times SU(3)_{C^\prime}$, 
see Fig.~\ref{Fig: model}. These new states do not couple to the Higgs even at the two loop 
level, so for fixed RG scale $\mu$ the two-loop potential is exactly the same as in the SM.
The potential is modified once we implement RG improvement and set $\mu = h$. 
The bifundamentals change the RG running of the QCD gauge coupling $g_3$ and 
the top quark Yukawa $y_t$:
\begin{align}
  \Delta \beta_{g_3} & = \frac{1}{(4\pi)^2}\left(\frac{2}{3} \Ncp g_3^3\right) 
+ \frac{1}{(4\pi)^4} \left(\frac{38}{3}\Ncp g_3^5\right), \\
\Delta \beta_{y_t} & = \frac{1}{(4\pi)^4}\left(\frac{40}{9}\Ncp g_3^4 y_t\right),
\end{align}
where $N_{c^\prime} = 3$ is number of colors in $SU(3)_{C^\prime}$.
As we evolve the coupling constants of the SM into the UV, 
these additional fermions slow the running of $g_3$; as a result $y_t$ runs 
faster toward zero due the larger $g_3$. The net effect is that the negative contribution of $y_t$ 
to the Higgs quartic $\lambda$ is reduced and it 
can run positive before the Planck scale. Note that this formally occurs at 
3 loops -- it is only captured by the leading and next-to-leading 
log resummation implemented by the RG improvement of the potential.
We demonstrate the effect of the bifundamentals with a 
mass scale of $100\;\TeV$ on $\lameff$ in Fig.~\ref{fig:lameff} as the solid red 
line. As before, the band around the solid line corresponds to $1\sigma$ variations of $M_t$.
Note that $\lameff$ runs positive at $h\sim 10^{16}\;\GeV$.

In the next section we study the extrema of the potential and 
determine confinement scale of QCD$^\prime$.

\subsection{Bifundamental Mass Scale}
\label{sec:bifunmass}

The scale at which the bifundamentals are integrated in or out is not a free parameter, 
but is rather fixed by the confinement scale of QCD$^\prime$, $\LamQCDp$.
The latter is determined by the masses of the $\bbZ_2$ quarks 
and the resulting RG evolution of $g_3^\prime$.

Tree-level mirror quark masses only depend on the mirror Higgs vev $h^\prime$ and 
their Yukawa couplings. The location of extrema of 
Eq.~\ref{Eq: eff pot} is obtained by solving 
\beq
\lameff(h^\prime) + \frac{1}{4}\beta_{\lameff}(h^\prime) = 0,
\label{eq:pot_der}
\eeq
where $\beta_{\lameff} = d \lameff/d\ln h'$.
Note that potential, and therefore $\lameff$ are gauge-dependent quantities. 
The vev $h^\prime$ inherits this gauge dependence and is therefore not physical 
-- see, e.g., Refs.~\cite{Patel:2011th,Andreassen:2014eha,Andreassen:2014gha} 
and references therein. Gauge invariant quantities 
can be obtained from pole masses~\cite{Hempfling:1994ar}. 
The Higgs expectation value shown in Fig.~\ref{Fig: hmin} is 
computed using the physical $W$ mass in the 
new minimum: $h_\mathrm{min} = 2 m_{W}/g$.
The Landau gauge self-energy needed for this is
available in Ref.~\cite{Martin:2015lxa}. 
In Fig.~\ref{Fig: hmin} we see that the Higgs expectation values 
can be large and one might worry that the following 
results can be significantly altered  by 
Planck-suppressed operators in the potential. We 
study these potential deformations in Appendix~\ref{sec:higherdim} and 
find that they do not change the main results of this section.

Once the mirror quark masses have been determined we run the QCD$^\prime$ coupling 
$g_3^\prime$ from the scale of the vev into the IR using the QCD beta function in the 
one, two, three or four loop approximation~\cite{vanRitbergen:1997va}. 
The scale at which the coupling becomes non-perturbative, $g_3^\prime\sim 4\pi$, 
defines the confinement scale $\LamQCDp$. 
As discussed in Sec.~\ref{sec:cpsol}, in the IR QCD$^\prime$ is a 
theory containing three massless flavors (the bifundamental quarks 
$\psi^{(c)}$), so its spectrum after confinement resembles low energy QCD. 
In particular, the physical scale of low energy 
QCD$^\prime$ corresponds to the mass of the lowest lying 
non-Goldstone meson -- the $\rho^\prime$. We determine the $\rho^\prime$ mass by
\beq
\frac{m_{\rho^\prime}}{m_\rho} = \frac{\LamQCDp^{(l)}}{\LamQCD^{(l)}},
\eeq
where $m_\rho = 0.77\;\GeV$ and $\LamQCD$ is the QCD confinement scale 
computed analogously to $\LamQCDp$. The superscript $l$ indicates that 
the ratio of scales depends on the loop order of the beta function used.
As in QCD, the lightest states are the Goldstone boson pions $\pi^\prime$ whose 
mass is determined by the explicit breaking of the global $SU(3)_L\times SU(3)_R$ 
through the gauging of QCD (see Sec.~\ref{sec:cpsol} and Ref.~\cite{Hook:2014cda})
\beq
m_{\pi^\prime}^2 \approx \frac{9\alpha_s(m_{\pi^\prime})}{4\pi} m_{\rho^\prime}^2.
\eeq

The dynamical scale $m_{\rho^\prime}$ roughly determines where the bifundamental states 
are integrated into the running of our $\alpha_s$ and the strongly coupled physics is integrated out. Thus 
it has an important effect on the running of the Higgs quartic $\lambda$. 
Evolution through the strongly-coupled threshold of QCD$^\prime$ 
can be done using dispersion relation methods. This problem is analogous to the 
running of $\alpha$ through the QCD confinement regime. 
In the SM the hadronic contribution to $\alpha(M_Z) - \alpha(0)$ is related to 
a measured cross-section using the optical theorem. We use the knowledge 
of low energy QCD to construct equivalent quantities in QCD$^\prime$ and 
evaluate $\alpha_s(\mu \gtrsim \mrhop) - \alpha_s(\mu\lesssim \mrhop)$. 
Below and above $\sim \mrhop$, $\alpha_s$ is evolved using the usual perturbative beta functions. 
The discontinuity takes into account the neglected RG effects of the $\pi'$ and
$\rho'$.  This procedure is described in detail in
Appendix~\ref{sec:strong_threshold}.

\begin{figure}[t]
  \centering
  \includegraphics{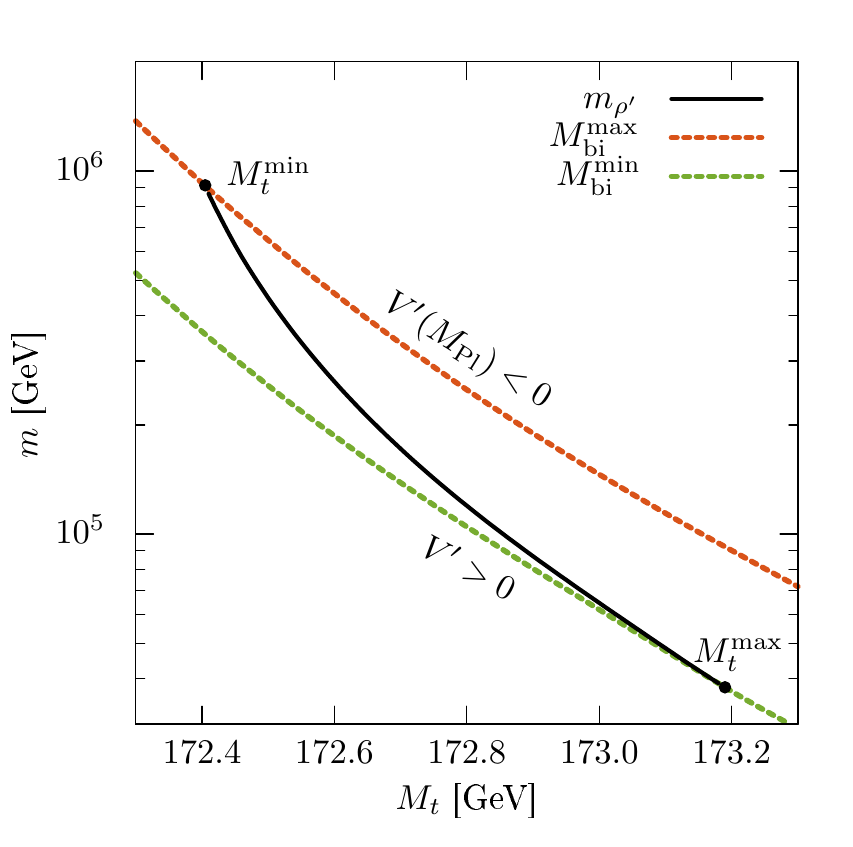}
  \caption{Minimum and maximum values of the bifundamental scale $\Mbi$ (dashed
  lines) as a function of the input top mass $M_t$. The solid line is the mass
of $\rho^\prime$ determined self-consistently from the confinement scale of
QCD$^\prime$ as described in the text. Below $\Mbi^\text{min}$ (lower dashed
line) no minimum in the potential is generated by loop corrections. Above $\Mbi^\text{max}$ (upper dashed line) the minimum occurs above $\Mpl$.}
  \label{fig:minmax_vs_mt}
\end{figure}

The confinement scale of QCD$^\prime$, $\LamQCDp$, determines when the 
the bifundamental states appear in the RGEs, which, in turn sets $h^\prime$, 
the vev in mirror sector. The latter feeds back into $\LamQCDp$ through 
the mirror quark masses. Thus $\LamQCDp$, or equivalently, $m_{\rho^\prime}$, 
must be determined recursively. A self-consistent solution for $\mrhop$ 
does not always exist. To understand this, consider 
fixing $M_t$ and integrating in the bifundamentals 
at a scale $\mu=M_{\mathrm{bi}}$. Above some maximum value of $\Mbi = \Mbi^\text{max}$, 
the RG effect of the bifundamentals is not enough to stabilize the potential 
at sub-Planckian field values, that is $V'(\Mpl) < 0$.
Similarly, below a minimum value of $\Mbi = \Mbi^\text{min}$ the 
stabilizing effect of the bifundamentals is
too large and the electroweak minimum is 
absolutely stable, i.e., Eq.~\ref{eq:pot_der} is never satisfied. 
The dependence of $\Mbi^\text{min}$ and $\Mbi^\text{max}$ on $M_t$ is shown in Fig.~\ref{fig:minmax_vs_mt} 
as the upper and lower dashed lines, respectively. 
For intermediate $\Mbi$, a second, non-electroweak minimum exists with a sub-Planckian vev, 
allowing us to search for a solution to 
\beq
\Mbi \sim \mrhop (\Mbi),
\label{eq:mrhop}
\eeq
where $\mrhop$ is determined as described above and the argument indicates that it 
is a functional of $\Mbi$ through the RG equations and potential minimization.  
Solutions to this equation exist only in a narrow range of $M_t$, as 
shown in Fig.~\ref{fig:minmax_vs_mt} by the solid black line. 
Above a maximum $M_t^{\mathrm{max}}$, the bifundamental scale needed 
to stabilize the potential becomes lower than the minimum allowed value (the lower dashed line). 
In the other limit, as $M_t$ is lowered the potential becomes more stable (due to smaller $y_t$)
and the $\mrhop$ needed to generate a new minimum increases, eventually crossing 
the maximum allowed value (upper dashed line) when $M_t = M_t^{\mathrm{min}}$.

We conclude this section by considering the sources of uncertainty in the physical outputs 
of this framework -- 
the mass scale of the bifundamental states and range of ``valid'' $M_t$. 
The dominant sources of this uncertainty are related to the 
strongly-coupled regime of QCD$^\prime$. 
First, the RG evolution through 
the strongly coupled regime of QCD$^\prime$ depends on the precise 
low energy spectrum of that theory. While we constructed a 
reasonable model based on symmetry arguments (see Appendix~\ref{sec:strong_threshold}), 
a non-perturbative determination of the spectrum and the resulting 
threshold correction to $\alpha_s$ is desirable. This can be achieved, 
for example, with lattice simulations or with holographic methods~\cite{Brodsky:2015oia}.
We estimated the uncertainty associated with this calculation by 
varying the matching scale at which this threshold is applied, as 
described in Appendix~\ref{sec:strong_threshold}. The resulting uncertainty bands are 
shown in Fig.~\ref{Fig: hmin}. This variation also shifts 
the valid range of $M_t$ by $\sim 0.5\;\GeV$, which is not 
shown in the figure.

Second, our determination 
of $\LamQCDp$ relies on perturbative beta functions, leading to different values 
depending on loop order used. The variation with loop order is shown in the 
right plot of Fig.~\ref{Fig: hmin}: the solid (dashed) line 
corresponds to the four (one) loop determination of $\LamQCDp$. 

Another uncertainty is associated with 
matching at the electroweak scale and variation of 
$\alpha_s(M_Z)$ and $M_h$ within their experimental 
error bars. In a full SM next-to-next-to-leading order (NNLO)
computation (the current state-of-the-art) this 
amounts to an uncertainty of $\sim 0.5\;\GeV$ on 
the critical value of $M_t$ that leads to 
a stable electroweak minimum~\cite{Bednyakov:2015sca,Iacobellis:2016eof}.
We expect a somewhat larger uncertainty on the postdicted range of $M_t$ 
with our approximations due to the absence of three-loop RGE.
Beyond NNLO, the neglected higher-order terms in the effective potential and 
beta functions are expected to have a negligible effect~\cite{Iacobellis:2016eof}.
Lastly, Planck suppressed operators can modify the shape of the potential at large 
field values. As long as the gauge invariant quantities associated with the vev 
are much smaller than $\Mpl$, we expect these (unknown) effects to be 
unimportant~\cite{Andreassen:2014eha}. Nevertheless, we consider 
the impact of these deformations in Appendix~\ref{sec:higherdim}, 
finding no qualitative changes to the main results of this section.

\section{Conclusion}
\label{sec:conclusion}
We have considered a solution to the hierarchy problem associated 
with a new scalar field. The new tree-level mass terms are related 
to the Standard Model Higgs mass by a discrete symmetry and are therefore protected 
from large corrections by the same mechanism that resolves the usual 
electroweak hierarchy problem. The new scalar obtains its mass and 
expectation value through dimensional transmutation of the Higgs quartic coupling, 
allowing these dimensionful quantities to be significantly different 
from the electroweak scale, while remaining technically natural.
We illustrated this mechanism in the context of an $\eta^\prime$ 
solution to the strong CP problem, which employs a $\bbZ_2$ symmetry and 
new massless quarks to rotate away the QCD theta angle. The resulting 
model predicts scalar color octet states and determines their 
mass in terms of the Standard Model input parameters (and 
the representation of the new quarks). 

The solution to the ``wrong'' hierarchy problem presented here, while extremely
predictive, leaves open a few questions.  First, what is the cosmological
history of these solutions? Aside from the constraint that the reheating
temperature is smaller than the scale at which parity or $\mathbb{Z}_2$ symmetry is
restored, there is the question of how the universe ended up in the
metastable state of one Higgs and in the true minimum of the other. 
It would be interesting to find a convincing reason for why this
could be a generic occurrence.

It is also interesting to determine if there exist solutions to other problems of the SM
which are testable in this framework. For example, one might utilize this
effect to solve the doublet triplet splitting problem in Grand Unified
Theories.  Because the other scalar obtains a expectation value, one would need to enlarge
$SU(3)_C$ to $SU(4)_C$ at high scales. Another use of this solution to the
hierarchy problem is to use a $\mathbb{Z}_2$ version of the currently excluded
Higgs as a flavon theory~\cite{Babu:1999me,Giudice:2008uua}.  Using the sum
$h^2 + h'^2$ as the flavon puts the flavor scale at a very safe but very
unobservable large value. This effect also fits well with Left-Right (LR)
symmetric models as it avoids the need for additional bifundamental
scalar fields (see e.g. Ref.~\cite{Senjanovic:1975rk}) leading to very
predictive but also unobservable scenarios.

Of course the most interesting extension would be applying this solution to
the SM Higgs boson. As in similar scenarios with 
Coleman-Weinberg electroweak symmetry breaking (see, e.g., Ref.~\cite{Hill:2014mqa}), 
this is difficult to achieve.
First, the Coleman-Weinberg minimum has a
Higgs mass parametrically smaller than its expectation value, something not
seen with the SM Higgs. Second, the cubic and quartic Higgs self-interactions would 
deviate significantly from their SM values. Third, the potential 
minimization condition, Eq.~\ref{eq:pot_der}, implies that at the radiatively-generated 
minimum the effective quartic has a positive beta function. 
For the SM Higgs boson the large top Yukawa drives this
beta function negative at low energies, so new bosonic degrees of freedom 
at relatively low masses are required to make it positive~\cite{Hill:2014mqa}.  

In this paper, another small step was taken towards finding non-standard
solutions to hierarchy problems.  These mechanisms have not been
thoroughly explored before and there may be many interesting solutions that
are still undiscovered.  In continuing along the direction of this work, it would be
extremely interesting if one could dimensionally transmute the Higgs
scale itself, from, e.g., the scale of the cosmological constant.

\section*{Acknowledgments}
We thank Stephen Martin, Michael Peskin, Lance Dixon, Brian Shuve 
for helpful discussions and David Morrissey for feedback on the manuscript. 
N.B. is supported by DOE Contract DE-AC02-76SF00515. 
A.H. is supported by the DOE Grant DE-SC0012012 and NSF Grant 1316699.

\appendix

\section{Integrating Through a Strongly Coupled Threshold\label{sec:strong_threshold}}

The RG evolution of the QCD coupling $g_3$ passes through 
the confinement scale of QCD$^\prime$. Above this threshold 
the degrees of freedom contributing to the beta function are the 
free quarks and the bifundamentals, while below only the quarks contribute. 
In the intermediate region, 
the degrees of freedom associated with the bifundamentals are strongly 
coupled and consist of various mirror hadrons that carry QCD charge. 
A similar situation arises in the SM when we want to evaluate the 
the shift in electromagnetic coupling $\Delta\alpha$ between momentum scales 
$q_0^2$ and $q^2$, with QCD confinement occurring inside this range. 
This is done using the once-subtracted dispersion relation for the gauge boson self-energy $\Pi(q^2)$~\cite{Actis:2010gg}
\beq
\Delta \alpha = \Pi(q^2)-\Pi(q_0^2) = \frac{(q^2-q_0^2)}{\pi}\int_{\sthr}^\infty ds \frac{\Im\Pi(s+i\epsilon)}{(s-q_0^2)(s-q^2)},
\label{eq:once_sub_dr}
\eeq
where $\sthr$ is the beginning of the branch cut corresponding 
to on-shell intermediate states, e.g. pions, $\rho$'s, etc. 
and $q_0^2 < \sthr$, i.e. $\Im \Pi(q_0^2) = 0$.
In our case the lightest intermediate states are the $\pi^\prime$.
Thus we take $\sthr = 4m_{\pi^\prime}^2$.

In order to evaluate the integrand in Eq.~\ref{eq:once_sub_dr} 
we consider the forward scattering amplitude $\bar q q \rightarrow \bar q q$. 
We can isolate the imaginary part of the gluon self-energy by considering only the 
processes that occur through an $s$-channel gluon exchange. There are others, 
such as box diagrams with two intermediate gluons, but these are higher order in 
$g_3$. Applying the optical theorem and summing over initial and final colors and spins 
we find 
\beq
\Im \Pi (s) = -
\left[
\frac{N^2_c}{(N^2_c-1)T_F}
\right]
\left[\frac{s}{4\pi \alpha_s}\right]
\sigma(\bar q q\rightarrow \text{hadrons}^\prime),
\eeq
where $N_c = 3$ and $T_F = 1/2$.
In the non-perturbative regime we must use the measured 
$\sigma(e^+ e^-\rightarrow \text{hadrons})$, 
suitably scaled to model 
$\sigma(\bar q q\rightarrow \text{hadrons}^\prime)$. 
Note, however, that the physical QCD spectrum \emph{is} dictated 
by explicit breaking of chiral symmetry through non-zero quark masses. 
This splits various states that would be degenerate in 
QCD$^\prime$. Thus a conservative approach 
is not to use the full $R(s)$ but only the contribution
from $e^+ e^-\rightarrow \pi^+\pi^-$ 
(corresponding to $\bar q q \rightarrow \pi^\prime\pi^\prime$ in QCD$^{\prime}$), which dominates 
at low energies. At higher energies we switch to the perturbative cross-section, where is 
it a good description of the data. 
By restricting ourselves to the $\pi\pi$ final state in the non-perturbative regime, the 
group-theoretic factors can be easily accounted for.
An analytical fit in terms masses of $\pi$ and the vector mesons for $e^+e^-\rightarrow \pi^+ \pi^-$ 
is given in Ref.~\cite{Lees:2012cj}. We use this result by rescaling the parameters of the fit 
to account for different meson masses in QCD$^\prime$.

Let us define a quantity $R^\prime (s)$ that resembles the SM hadronic $R$ function:
\beq
R^\prime (s) = \kappa\frac{\sigma(\bar q q\rightarrow \pi^\prime\pi^\prime)}
{\sigma(\bar q q\rightarrow \bar{q}^\prime q^\prime)} = -\frac{3\kappa}{\alpha_s T_F} \Im \Pi(s),
\label{eq:rp_def}
\eeq
where
\beq
\sigma(\bar q q\rightarrow \bar{q}^\prime q^\prime)
= 
\left[
\frac{(N^2_c-1)T_F^2}{N^2_c}
\right]
\frac{4\pi \alpha_s^2}{3s},
\label{eq:sigqq}
\eeq
and $\kappa$ is a numerical factor to be fixed later.
Using Eq.~\ref{eq:rp_def} in Eq.~\ref{eq:once_sub_dr}  we find 
\beq
\Delta \alpha_s^{(\mathrm{QCD}^\prime)} = \Pi(q^2)- \Pi(q_0^2) = -\frac{\alpha_s T_F (q^2-q_0^2)}{3\pi\kappa}\int_{\sthr}^\infty ds \frac{R^\prime(s)}{(s-q_0^2)(s-q^2)},
\label{eq:delta_pi_hadronic}
\eeq
where the superscript indicates that this is only the contribution of the QCD$^\prime$ states - running of $\alpha_s$ due to SM quarks still needs to be included.
This is identical to the usual formula for $\Delta \alpha_{\mathrm{had}}$~\cite{Actis:2010gg} for 
$q_0^2 = 0$, and $\kappa,\;T_F\rightarrow 1$.

The integral in Eq.~\ref{eq:delta_pi_hadronic} is performed numerically using 
a data-driven model for $R^\prime(s)$.
We use the observed $R$ to model $R^\prime$ in the non-perturbative regime. 
We take the analytic model of the $\pi\pi$ contribution to $R$ from Ref.~\cite{Lees:2012cj} and 
replace all dimensionful parameters by their QCD$^\prime$ analogues and 
find
\beq
R \approx\frac{\sigma(e^+ e^-\rightarrow \pi^+\pi^-)}{\sigma(e^+ e^-\rightarrow \mu^+\mu^-)}
= \frac{T_F}{C_A}\frac{\sigma(\bar q q\rightarrow \pi^\prime\pi^\prime)}{\sigma (\bar q q \rightarrow \bar q^\prime q\prime)}
= R^\prime,
\label{eq:rprime_def}
\eeq
where $C_A=N_c$. This defines
\beq
\kappa = \frac{T_F}{C_A}.
\eeq
Equation~\ref{eq:rprime_def} must be used in the non-perturbative regime. 
We switch to the perturbative result for annihilation into the massless bifundamentals
\beq
R^\prime (s) \approx \kappa N_c,
\eeq
when the perturbative and non-perturbative results become equal above $s \gtrsim (m_{\rho^\prime})^2$.
The threshold correction of Eq.~\ref{eq:delta_pi_hadronic} is 
applied between $q_0^2 \approx (m_{\pi^\prime})^2$ and 
$q^2 \approx (3 m_{\rho^\prime})^2$, above which normal perturbative RG
evolution is resumed. The values of $q_0^2$ and $q^2$ are chosen to lie well
below and well above the hadronic resonances, respectively.
The uncertainty bands in Fig.~\ref{Fig: hmin} were estimated by varying 
$q$ from $3 m_{\rho^\prime}$ to $6 m_{\rho^\prime}$.

\section{Higher Dimensional Operators in the Scalar Potential}
\label{sec:higherdim}
Stability constraints in the SM are sensitive to higher-dimensional 
operators in the potential, even if they are suppressed by $\Mpl$~\cite{Andreassen:2014gha}. 
They can also drastically alter the lifetime of metastable vacua~\cite{Branchina:2013jra,Andreassen:2016cvx}.
In this Appendix we investigate the sensitivity of our results 
to the presence of operators of the form $\pm |H|^6/\Mpl^2$. 
For $h\sim 10^{17}\;\GeV$ these terms are comparable in magnitude to the 
other terms in the effective potential; comparing this 
with Fig.~\ref{Fig: hmin} we see that such operators can 
have a significant effect for some values of $M_t$. 

\begin{figure}
\centering
\includegraphics[width=0.47\textwidth]{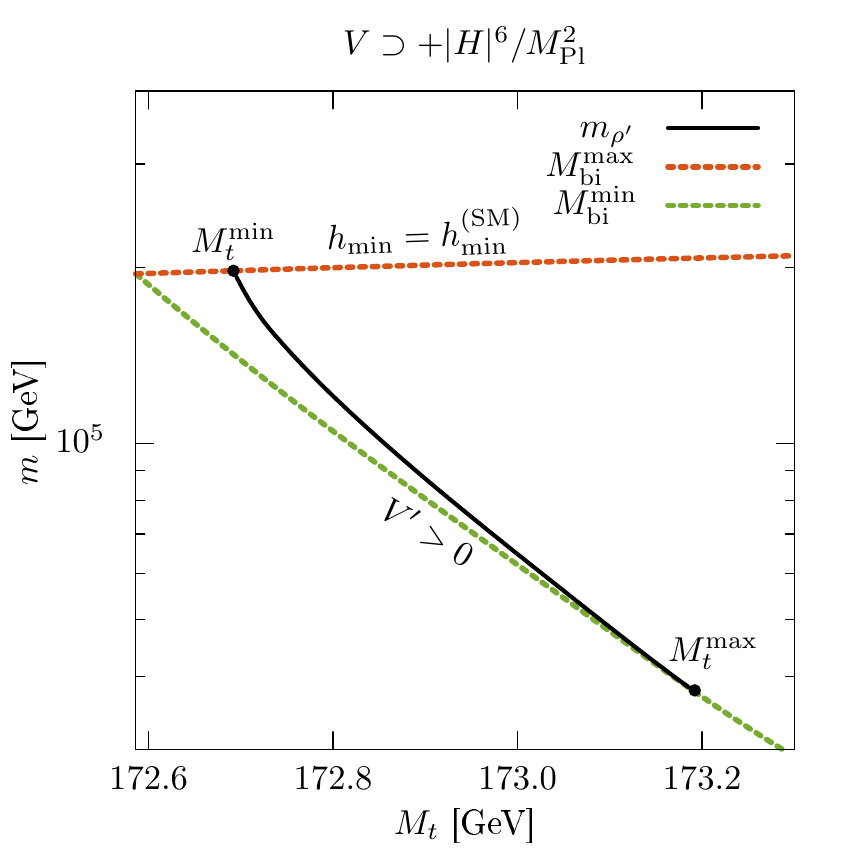}
\includegraphics[width=0.47\textwidth]{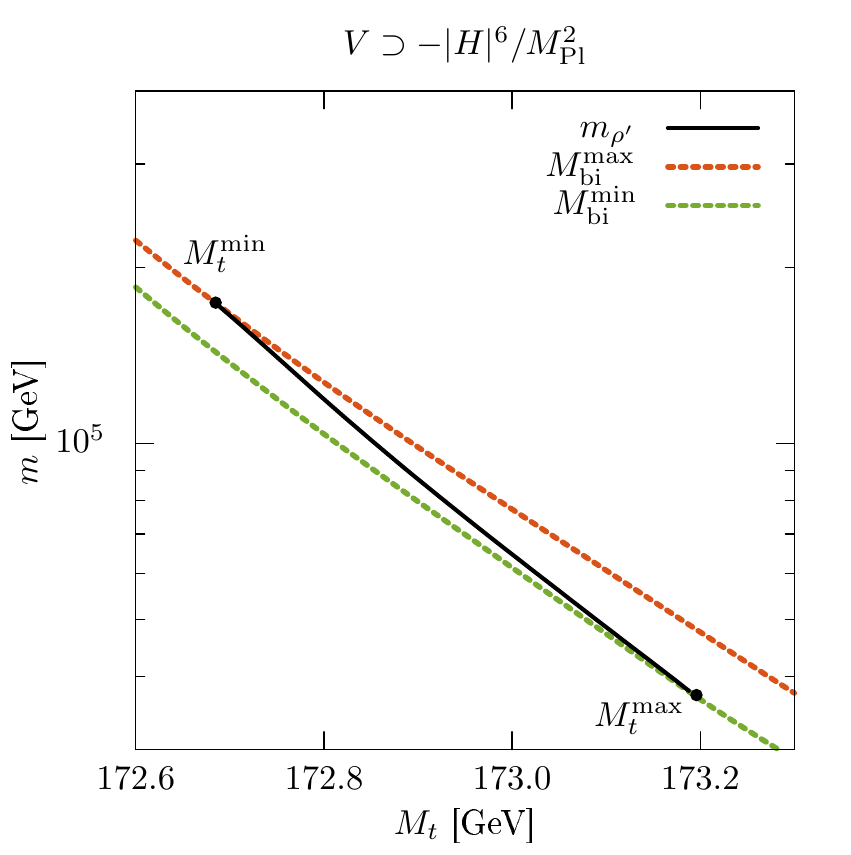}
\caption{Prediction of $m_{\rho^\prime}$ and viable range of $M_t$ in 
the presence of Planck-suppressed operators $\pm |H|^6/\Mpl^2$. 
The minimum and maximum values of the bifundamental scale are 
defined in the text.\label{fig:higher_dim_op}
}
\end{figure}

First, we consider the operators $+|H|^6/\Mpl^2$. This term stabilizes the potential 
at large field values. Even without additional matter the Higgs potential 
develops a minimum with expectation value $h_{\mathrm{min}}^{(\mathrm{SM})}\sim 10^{18}\;\GeV$.
This vev can be used to define the maximum bifundamental mass scale $M_{\mathrm{bi}}^{\mathrm{max}}$, 
since the bifundamental states give an additional stabilizing effect 
leading to $h_{\mathrm{min}} < h_{\mathrm{min}}^{(\mathrm{SM})}$. 
The minimum scale $M_{\mathrm{bi}}^{\mathrm{min}}$ is defined as 
in Sec.~\ref{sec:bifunmass}. With these bounds in hand, we solve 
the self-consistency equation, Eq.~\ref{eq:mrhop}, for $m_{\rho^\prime}$.
The resulting solution is shown in the left plot of Fig.~\ref{fig:higher_dim_op}.
As before the solution only exists in a small range of $M_t$. 

Next we consider $-|H|^6/\Mpl^2$. Since this destabilizes the potential, it 
must be compensated by even higher dimensional operators like $|H|^8/\Mpl^2$. 
Interestingly, even in this case the bifundamental states can lead to 
a new minimum in the potential. The bounds on the bifundamental mass 
scale arise from considerations similar to those in Sec.~\ref{sec:bifunmass}. 
Above a maximum value  $M_{\mathrm{bi}}^{\mathrm{max}}$, no second minimum exists.
Below the minimum $M_{\mathrm{bi}}^{\mathrm{min}}$ the potential develops an instability, but only 
due to the higher dimensional operator and not due to $y_t$. 
This gives rise to very tight band of bifundamental mass scales in 
which a solution to Eq.~\ref{eq:mrhop} may exist. This is shown 
in the right plot of Fig.~\ref{fig:higher_dim_op}.

In both cases above, the additional Planck-suppressed operators 
reduce the range of $M_t$ for which a self-consistent solution for $m_{\rho'}$ exists.
Comparing Figs.~\ref{fig:minmax_vs_mt} and~\ref{fig:higher_dim_op} we find 
that the resulting values of $m_{\rho'}$ are not very different 
from the case with no higher-dimensional operators.
Thus we conclude that these deformations of the potential do not affect the qualitative 
features of the mechanism presented in Sec.~\ref{sec:dimtrans}.

\bibliographystyle{utphys}
\bibliography{ref}

\providecommand{\href}[2]{#2}\begingroup\raggedright\begin{thebibliography}{10}

\bibitem{Graham:2015cka}
P.~W. Graham, D.~E. Kaplan, and S.~Rajendran, ``{Cosmological Relaxation of the
  Electroweak Scale},''
  \href{http://dx.doi.org/10.1103/PhysRevLett.115.221801}{{\em Phys. Rev.
  Lett.} {\bf 115} (2015) no.~22, 221801},
\href{http://arxiv.org/abs/1504.07551}{{\tt arXiv:1504.07551 [hep-ph]}}.

\bibitem{God_knows_when}
N.~Arkani-Hamed, T.~Cohen, R.~D'Agnolo, A.~Hook, H.~D. Kim, and D.~Pinner,
  ``{NNaturalness},'' {\em To appear}  .

\bibitem{Coleman:1973jx}
S.~R. Coleman and E.~J. Weinberg, ``{Radiative Corrections as the Origin of
  Spontaneous Symmetry Breaking},''
\href{http://dx.doi.org/10.1103/PhysRevD.7.1888}{{\em Phys. Rev.} {\bf D7}
  (1973)  1888--1910}.

\bibitem{Degrassi:2012ry}
G.~Degrassi, S.~Di~Vita, J.~Elias-Miro, J.~R. Espinosa, G.~F. Giudice,
  G.~Isidori, and A.~Strumia, ``{Higgs mass and vacuum stability in the
  Standard Model at NNLO},''
  \href{http://dx.doi.org/10.1007/JHEP08(2012)098}{{\em JHEP} {\bf 08} (2012)
  098},
\href{http://arxiv.org/abs/1205.6497}{{\tt arXiv:1205.6497 [hep-ph]}}.

\bibitem{Buttazzo:2013uya}
D.~Buttazzo, G.~Degrassi, P.~P. Giardino, G.~F. Giudice, F.~Sala, A.~Salvio,
  and A.~Strumia, ``{Investigating the near-criticality of the Higgs boson},''
  \href{http://dx.doi.org/10.1007/JHEP12(2013)089}{{\em JHEP} {\bf 12} (2013)
  089},
\href{http://arxiv.org/abs/1307.3536}{{\tt arXiv:1307.3536 [hep-ph]}}.

\bibitem{Bednyakov:2015sca}
A.~V. Bednyakov, B.~A. Kniehl, A.~F. Pikelner, and O.~L. Veretin, ``{Stability
  of the Electroweak Vacuum: Gauge Independence and Advanced Precision},''
  \href{http://dx.doi.org/10.1103/PhysRevLett.115.201802}{{\em Phys. Rev.
  Lett.} {\bf 115} (2015) no.~20, 201802},
\href{http://arxiv.org/abs/1507.08833}{{\tt arXiv:1507.08833 [hep-ph]}}.

\bibitem{Einhorn:2007rv}
M.~B. Einhorn and D.~R.~T. Jones, ``{The Effective potential, the
  renormalisation group and vacuum stability},''
  \href{http://dx.doi.org/10.1088/1126-6708/2007/04/051}{{\em JHEP} {\bf 04}
  (2007)  051},
\href{http://arxiv.org/abs/hep-ph/0702295}{{\tt arXiv:hep-ph/0702295
  [HEP-PH]}}.

\bibitem{Salvio:2014soa}
A.~Salvio and A.~Strumia, ``{Agravity},''
  \href{http://dx.doi.org/10.1007/JHEP06(2014)080}{{\em JHEP} {\bf 06} (2014)
  080},
\href{http://arxiv.org/abs/1403.4226}{{\tt arXiv:1403.4226 [hep-ph]}}.

\bibitem{Okun:2006eb}
L.~B. Okun, ``{Mirror particles and mirror matter: 50 years of speculations and
  search},'' \href{http://dx.doi.org/10.1070/PU2007v050n04ABEH006227}{{\em
  Phys. Usp.} {\bf 50} (2007)  380--389},
\href{http://arxiv.org/abs/hep-ph/0606202}{{\tt arXiv:hep-ph/0606202
  [hep-ph]}}.

\bibitem{Chacko:2005pe}
Z.~Chacko, H.-S. Goh, and R.~Harnik, ``{The Twin Higgs: Natural electroweak
  breaking from mirror symmetry},''
  \href{http://dx.doi.org/10.1103/PhysRevLett.96.231802}{{\em Phys. Rev. Lett.}
  {\bf 96} (2006)  231802},
\href{http://arxiv.org/abs/hep-ph/0506256}{{\tt arXiv:hep-ph/0506256
  [hep-ph]}}.

\bibitem{An:2009vq}
H.~An, S.-L. Chen, R.~N. Mohapatra, and Y.~Zhang, ``{Leptogenesis as a Common
  Origin for Matter and Dark Matter},''
  \href{http://dx.doi.org/10.1007/JHEP03(2010)124}{{\em JHEP} {\bf 03} (2010)
  124},
\href{http://arxiv.org/abs/0911.4463}{{\tt arXiv:0911.4463 [hep-ph]}}.

\bibitem{Berezhiani:2000gh}
Z.~Berezhiani, L.~Gianfagna, and M.~Giannotti, ``{Strong CP problem and mirror
  world: The Weinberg-Wilczek axion revisited},''
  \href{http://dx.doi.org/10.1016/S0370-2693(00)01392-7}{{\em Phys. Lett.} {\bf
  B500} (2001)  286--296},
\href{http://arxiv.org/abs/hep-ph/0009290}{{\tt arXiv:hep-ph/0009290
  [hep-ph]}}.

\bibitem{Hook:2014cda}
A.~Hook, ``{Anomalous solutions to the strong CP problem},''
  \href{http://dx.doi.org/10.1103/PhysRevLett.114.141801}{{\em Phys. Rev.
  Lett.} {\bf 114} (2015) no.~14, 141801},
\href{http://arxiv.org/abs/1411.3325}{{\tt arXiv:1411.3325 [hep-ph]}}.

\bibitem{D'Agnolo:2015uta}
R.~T. D'Agnolo and A.~Hook, ``{Finding the Strong CP problem at the LHC},''
\href{http://arxiv.org/abs/1507.00336}{{\tt arXiv:1507.00336 [hep-ph]}}.

\bibitem{Foot:2014mia}
R.~Foot, ``{Mirror dark matter: Cosmology, galaxy structure and direct
  detection},'' \href{http://dx.doi.org/10.1142/S0217751X14300130}{{\em Int. J.
  Mod. Phys.} {\bf A29} (2014)  1430013},
\href{http://arxiv.org/abs/1401.3965}{{\tt arXiv:1401.3965 [astro-ph.CO]}}.

\bibitem{Albaid:2015axa}
A.~Albaid, M.~Dine, and P.~Draper, ``{Strong CP and SUZ$_{2}$},''
  \href{http://dx.doi.org/10.1007/JHEP12(2015)046}{{\em JHEP} {\bf 12} (2015)
  046},
\href{http://arxiv.org/abs/1510.03392}{{\tt arXiv:1510.03392 [hep-ph]}}.

\bibitem{Dine:2015jga}
M.~Dine and P.~Draper, ``{Challenges for the Nelson-Barr Mechanism},''
  \href{http://dx.doi.org/10.1007/JHEP08(2015)132}{{\em JHEP} {\bf 08} (2015)
  132},
\href{http://arxiv.org/abs/1506.05433}{{\tt arXiv:1506.05433 [hep-ph]}}.

\bibitem{Ellis:1978hq}
J.~R. Ellis and M.~K. Gaillard, ``{Strong and Weak CP Violation},''
\href{http://dx.doi.org/10.1016/0550-3213(79)90297-9}{{\em Nucl.Phys.} {\bf
  B150} (1979)  141}.

\bibitem{Khriplovich:1993pf}
I.~B. Khriplovich and A.~I. Vainshtein, ``{Infinite renormalization of Theta
  term and Jarlskog invariant for CP violation},''
  \href{http://dx.doi.org/10.1016/0550-3213(94)90419-7}{{\em Nucl. Phys.} {\bf
  B414} (1994)  27--32},
\href{http://arxiv.org/abs/hep-ph/9308334}{{\tt arXiv:hep-ph/9308334
  [hep-ph]}}.

\bibitem{Agashe:2014kda}
{\bf Particle Data Group} Collaboration, K.~A. Olive {\em et al.}, ``{Review of
  Particle Physics},''
\href{http://dx.doi.org/10.1088/1674-1137/38/9/090001}{{\em Chin. Phys.} {\bf
  C38} (2014)  090001}.

\bibitem{Fujii:2015jha}
K.~Fujii {\em et al.}, ``{Physics Case for the International Linear
  Collider},''
\href{http://arxiv.org/abs/1506.05992}{{\tt arXiv:1506.05992 [hep-ex]}}.

\bibitem{Han:2010rf}
T.~Han, I.~Lewis, and Z.~Liu, ``{Colored Resonant Signals at the LHC: Largest
  Rate and Simplest Topology},''
  \href{http://dx.doi.org/10.1007/JHEP12(2010)085}{{\em JHEP} {\bf 12} (2010)
  085},
\href{http://arxiv.org/abs/1010.4309}{{\tt arXiv:1010.4309 [hep-ph]}}.

\bibitem{Chen:2014haa}
C.-Y. Chen, A.~Freitas, T.~Han, and K.~S.~M. Lee, ``{Heavy Color-Octet
  Particles at the LHC},''
  \href{http://dx.doi.org/10.1007/JHEP05(2015)135}{{\em JHEP} {\bf 05} (2015)
  135},
\href{http://arxiv.org/abs/1410.8113}{{\tt arXiv:1410.8113 [hep-ph]}}.

\bibitem{Cacciapaglia:2015eqa}
G.~Cacciapaglia, H.~Cai, A.~Deandrea, T.~Flacke, S.~J. Lee, and A.~Parolini,
  ``{Composite scalars at the LHC: the Higgs, the Sextet and the Octet},''
  \href{http://dx.doi.org/10.1007/JHEP11(2015)201}{{\em JHEP} {\bf 11} (2015)
  201},
\href{http://arxiv.org/abs/1507.02283}{{\tt arXiv:1507.02283 [hep-ph]}}.

\bibitem{Arkani-Hamed:2015vfh}
N.~Arkani-Hamed, T.~Han, M.~Mangano, and L.-T. Wang, ``{Physics Opportunities
  of a 100 TeV Proton-Proton Collider},''
\href{http://arxiv.org/abs/1511.06495}{{\tt arXiv:1511.06495 [hep-ph]}}.

\bibitem{Kamionkowski:1992mf}
M.~Kamionkowski and J.~March-Russell, ``{Planck scale physics and the
  Peccei-Quinn mechanism},''
  \href{http://dx.doi.org/10.1016/0370-2693(92)90492-M}{{\em Phys. Lett.} {\bf
  B282} (1992)  137--141},
\href{http://arxiv.org/abs/hep-th/9202003}{{\tt arXiv:hep-th/9202003
  [hep-th]}}.

\bibitem{Baker:2006ts}
C.~A. Baker {\em et al.}, ``{An Improved experimental limit on the electric
  dipole moment of the neutron},''
  \href{http://dx.doi.org/10.1103/PhysRevLett.97.131801}{{\em Phys. Rev. Lett.}
  {\bf 97} (2006)  131801},
\href{http://arxiv.org/abs/hep-ex/0602020}{{\tt arXiv:hep-ex/0602020
  [hep-ex]}}.

\bibitem{Griffith:2009zz}
W.~C. Griffith, M.~D. Swallows, T.~H. Loftus, M.~V. Romalis, B.~R. Heckel, and
  E.~N. Fortson, ``{Improved Limit on the Permanent Electric Dipole Moment of
  Hg-199},''
\href{http://dx.doi.org/10.1103/PhysRevLett.102.101601}{{\em Phys. Rev. Lett.}
  {\bf 102} (2009)  101601}.

\bibitem{Engel:2013lsa}
J.~Engel, M.~J. Ramsey-Musolf, and U.~van Kolck, ``{Electric Dipole Moments of
  Nucleons, Nuclei, and Atoms: The Standard Model and Beyond},''
  \href{http://dx.doi.org/10.1016/j.ppnp.2013.03.003}{{\em Prog. Part. Nucl.
  Phys.} {\bf 71} (2013)  21--74},
\href{http://arxiv.org/abs/1303.2371}{{\tt arXiv:1303.2371 [nucl-th]}}.

\bibitem{Akhmedov:1992hh}
E.~K. Akhmedov, Z.~G. Berezhiani, and G.~Senjanovic, ``{Planck scale physics
  and neutrino masses},''
  \href{http://dx.doi.org/10.1103/PhysRevLett.69.3013}{{\em Phys. Rev. Lett.}
  {\bf 69} (1992)  3013--3016},
\href{http://arxiv.org/abs/hep-ph/9205230}{{\tt arXiv:hep-ph/9205230
  [hep-ph]}}.

\bibitem{Andreassen:2014gha}
A.~Andreassen, W.~Frost, and M.~D. Schwartz, ``{Consistent Use of the Standard
  Model Effective Potential},''
  \href{http://dx.doi.org/10.1103/PhysRevLett.113.241801}{{\em Phys. Rev.
  Lett.} {\bf 113} (2014) no.~24, 241801},
\href{http://arxiv.org/abs/1408.0292}{{\tt arXiv:1408.0292 [hep-ph]}}.

\bibitem{Chao:2012mx}
W.~Chao, M.~Gonderinger, and M.~J. Ramsey-Musolf, ``{Higgs Vacuum Stability,
  Neutrino Mass, and Dark Matter},''
  \href{http://dx.doi.org/10.1103/PhysRevD.86.113017}{{\em Phys. Rev.} {\bf
  D86} (2012)  113017},
\href{http://arxiv.org/abs/1210.0491}{{\tt arXiv:1210.0491 [hep-ph]}}.

\bibitem{Chen:2012faa}
C.-S. Chen and Y.~Tang, ``{Vacuum stability, neutrinos, and dark matter},''
  \href{http://dx.doi.org/10.1007/JHEP04(2012)019}{{\em JHEP} {\bf 04} (2012)
  019},
\href{http://arxiv.org/abs/1202.5717}{{\tt arXiv:1202.5717 [hep-ph]}}.

\bibitem{EliasMiro:2012ay}
J.~Elias-Miro, J.~R. Espinosa, G.~F. Giudice, H.~M. Lee, and A.~Strumia,
  ``{Stabilization of the Electroweak Vacuum by a Scalar Threshold Effect},''
  \href{http://dx.doi.org/10.1007/JHEP06(2012)031}{{\em JHEP} {\bf 06} (2012)
  031},
\href{http://arxiv.org/abs/1203.0237}{{\tt arXiv:1203.0237 [hep-ph]}}.

\bibitem{Lebedev:2012zw}
O.~Lebedev, ``{On Stability of the Electroweak Vacuum and the Higgs Portal},''
  \href{http://dx.doi.org/10.1140/epjc/s10052-012-2058-2}{{\em Eur. Phys. J.}
  {\bf C72} (2012)  2058},
\href{http://arxiv.org/abs/1203.0156}{{\tt arXiv:1203.0156 [hep-ph]}}.

\bibitem{Espinosa:2015kwx}
J.~R. Espinosa, ``{Implications of the top (and Higgs) mass for vacuum
  stability},'' in {\em {8th International Workshop on Top Quark Physics
  (TOP2015) Ischia, NA, Italy, September 14-18, 2015}}.
\newblock 2015.
\newblock \href{http://arxiv.org/abs/1512.01222}{{\tt arXiv:1512.01222
  [hep-ph]}}.
\newblock
\url{http://inspirehep.net/record/1407977/files/arXiv:1512.01222.pdf}.
\newblock

\bibitem{Kastening:1991gv}
B.~M. Kastening, ``{Renormalization group improvement of the effective
  potential in massive phi**4 theory},''
\href{http://dx.doi.org/10.1016/0370-2693(92)90021-U}{{\em Phys. Lett.} {\bf
  B283} (1992)  287--292}.

\bibitem{Bando:1992np}
M.~Bando, T.~Kugo, N.~Maekawa, and H.~Nakano, ``{Improving the effective
  potential},'' \href{http://dx.doi.org/10.1016/0370-2693(93)90725-W}{{\em
  Phys. Lett.} {\bf B301} (1993)  83--89},
\href{http://arxiv.org/abs/hep-ph/9210228}{{\tt arXiv:hep-ph/9210228
  [hep-ph]}}.

\bibitem{Casas:1994qy}
J.~A. Casas, J.~R. Espinosa, and M.~Quiros, ``{Improved Higgs mass stability
  bound in the standard model and implications for supersymmetry},''
  \href{http://dx.doi.org/10.1016/0370-2693(94)01404-Z}{{\em Phys. Lett.} {\bf
  B342} (1995)  171--179},
\href{http://arxiv.org/abs/hep-ph/9409458}{{\tt arXiv:hep-ph/9409458
  [hep-ph]}}.

\bibitem{Ford:1992pn}
C.~Ford, I.~Jack, and D.~R.~T. Jones, ``{The Standard model effective potential
  at two loops},'' \href{http://dx.doi.org/10.1016/0550-3213(92)90165-8}{{\em
  Nucl. Phys.} {\bf B387} (1992)  373--390},
  \href{http://arxiv.org/abs/hep-ph/0111190}{{\tt arXiv:hep-ph/0111190
  [hep-ph]}}.
[Erratum: Nucl. Phys.B504,551(1997)].

\bibitem{Martin:2001vx}
S.~P. Martin, ``{Two loop effective potential for a general renormalizable
  theory and softly broken supersymmetry},''
  \href{http://dx.doi.org/10.1103/PhysRevD.65.116003}{{\em Phys. Rev.} {\bf
  D65} (2002)  116003},
\href{http://arxiv.org/abs/hep-ph/0111209}{{\tt arXiv:hep-ph/0111209
  [hep-ph]}}.

\bibitem{Martin:2014cxa}
S.~P. Martin and D.~G. Robertson, ``{Higgs boson mass in the Standard Model at
  two-loop order and beyond},''
  \href{http://dx.doi.org/10.1103/PhysRevD.90.073010}{{\em Phys. Rev.} {\bf
  D90} (2014) no.~7, 073010},
\href{http://arxiv.org/abs/1407.4336}{{\tt arXiv:1407.4336 [hep-ph]}}.

\bibitem{Machacek:1983tz}
M.~E. Machacek and M.~T. Vaughn, ``{Two Loop Renormalization Group Equations in
  a General Quantum Field Theory. 1. Wave Function Renormalization},''
\href{http://dx.doi.org/10.1016/0550-3213(83)90610-7}{{\em Nucl.Phys.} {\bf
  B222} (1983)  83}.

\bibitem{Machacek:1983fi}
M.~E. Machacek and M.~T. Vaughn, ``{Two Loop Renormalization Group Equations in
  a General Quantum Field Theory. 2. Yukawa Couplings},''
\href{http://dx.doi.org/10.1016/0550-3213(84)90533-9}{{\em Nucl.Phys.} {\bf
  B236} (1984)  221}.

\bibitem{Machacek:1984zw}
M.~E. Machacek and M.~T. Vaughn, ``{Two Loop Renormalization Group Equations in
  a General Quantum Field Theory. 3. Scalar Quartic Couplings},''
\href{http://dx.doi.org/10.1016/0550-3213(85)90040-9}{{\em Nucl.Phys.} {\bf
  B249} (1985)  70}.

\bibitem{Luo:2002ti}
M.-x. Luo, H.-w. Wang, and Y.~Xiao, ``{Two loop renormalization group equations
  in general gauge field theories},''
  \href{http://dx.doi.org/10.1103/PhysRevD.67.065019}{{\em Phys.Rev.} {\bf D67}
  (2003)  065019},
\href{http://arxiv.org/abs/hep-ph/0211440}{{\tt arXiv:hep-ph/0211440
  [hep-ph]}}.

\bibitem{Staub:2013tta}
F.~Staub, ``{SARAH 4: A tool for (not only SUSY) model builders},''
  \href{http://dx.doi.org/10.1016/j.cpc.2014.02.018}{{\em Comput.Phys.Commun.}
  {\bf 185} (2014)  1773--1790},
\href{http://arxiv.org/abs/1309.7223}{{\tt arXiv:1309.7223 [hep-ph]}}.

\bibitem{Lyonnet:2013dna}
F.~Lyonnet, I.~Schienbein, F.~Staub, and A.~Wingerter, ``{PyR@TE:
  Renormalization Group Equations for General Gauge Theories},''
  \href{http://dx.doi.org/10.1016/j.cpc.2013.12.002}{{\em Comput. Phys.
  Commun.} {\bf 185} (2014)  1130--1152},
\href{http://arxiv.org/abs/1309.7030}{{\tt arXiv:1309.7030 [hep-ph]}}.

\bibitem{Patel:2011th}
H.~H. Patel and M.~J. Ramsey-Musolf, ``{Baryon Washout, Electroweak Phase
  Transition, and Perturbation Theory},''
  \href{http://dx.doi.org/10.1007/JHEP07(2011)029}{{\em JHEP} {\bf 07} (2011)
  029},
\href{http://arxiv.org/abs/1101.4665}{{\tt arXiv:1101.4665 [hep-ph]}}.

\bibitem{Andreassen:2014eha}
A.~Andreassen, W.~Frost, and M.~D. Schwartz, ``{Consistent Use of Effective
  Potentials},'' \href{http://dx.doi.org/10.1103/PhysRevD.91.016009}{{\em Phys.
  Rev.} {\bf D91} (2015) no.~1, 016009},
\href{http://arxiv.org/abs/1408.0287}{{\tt arXiv:1408.0287 [hep-ph]}}.

\bibitem{Hempfling:1994ar}
R.~Hempfling and B.~A. Kniehl, ``{On the relation between the fermion pole mass
  and MS Yukawa coupling in the standard model},''
  \href{http://dx.doi.org/10.1103/PhysRevD.51.1386}{{\em Phys. Rev.} {\bf D51}
  (1995)  1386--1394},
\href{http://arxiv.org/abs/hep-ph/9408313}{{\tt arXiv:hep-ph/9408313
  [hep-ph]}}.

\bibitem{Martin:2015lxa}
S.~P. Martin, ``{Pole mass of the W boson at two-loop order in the pure
  $\overline {MS}$ scheme},''
  \href{http://dx.doi.org/10.1103/PhysRevD.91.114003}{{\em Phys. Rev.} {\bf
  D91} (2015) no.~11, 114003},
\href{http://arxiv.org/abs/1503.03782}{{\tt arXiv:1503.03782 [hep-ph]}}.

\bibitem{vanRitbergen:1997va}
T.~van Ritbergen, J.~A.~M. Vermaseren, and S.~A. Larin, ``{The Four loop beta
  function in quantum chromodynamics},''
  \href{http://dx.doi.org/10.1016/S0370-2693(97)00370-5}{{\em Phys. Lett.} {\bf
  B400} (1997)  379--384},
\href{http://arxiv.org/abs/hep-ph/9701390}{{\tt arXiv:hep-ph/9701390
  [hep-ph]}}.

\bibitem{Brodsky:2015oia}
S.~J. Brodsky, A.~Deur, G.~F. de~Téramond, and H.~G. Dosch, ``{Light-Front
  Holography and Superconformal Quantum Mechanics: A New Approach to Hadron
  Structure and Color Confinement},''
  \href{http://dx.doi.org/10.1142/S2010194515600812}{{\em Int. J. Mod. Phys.
  Conf. Ser.} {\bf 39} (2015)  1560081},
\href{http://arxiv.org/abs/1510.01011}{{\tt arXiv:1510.01011 [hep-ph]}}.

\bibitem{Iacobellis:2016eof}
G.~Iacobellis and I.~Masina, ``{Stationary configurations of the Standard Model
  Higgs potential: electroweak stability and rising inflection point},''
\href{http://arxiv.org/abs/1604.06046}{{\tt arXiv:1604.06046 [hep-ph]}}.

\bibitem{Babu:1999me}
K.~S. Babu and S.~Nandi, ``{Natural fermion mass hierarchy and new signals for
  the Higgs boson},'' \href{http://dx.doi.org/10.1103/PhysRevD.62.033002}{{\em
  Phys. Rev.} {\bf D62} (2000)  033002},
\href{http://arxiv.org/abs/hep-ph/9907213}{{\tt arXiv:hep-ph/9907213
  [hep-ph]}}.

\bibitem{Giudice:2008uua}
G.~F. Giudice and O.~Lebedev, ``{Higgs-dependent Yukawa couplings},''
  \href{http://dx.doi.org/10.1016/j.physletb.2008.05.062}{{\em Phys. Lett.}
  {\bf B665} (2008)  79--85},
\href{http://arxiv.org/abs/0804.1753}{{\tt arXiv:0804.1753 [hep-ph]}}.

\bibitem{Senjanovic:1975rk}
G.~Senjanovic and R.~N. Mohapatra, ``{Exact Left-Right Symmetry and Spontaneous
  Violation of Parity},''
\href{http://dx.doi.org/10.1103/PhysRevD.12.1502}{{\em Phys. Rev.} {\bf D12}
  (1975)  1502}.

\bibitem{Hill:2014mqa}
C.~T. Hill, ``{Is the Higgs Boson Associated with Coleman-Weinberg Dynamical
  Symmetry Breaking?},''
  \href{http://dx.doi.org/10.1103/PhysRevD.89.073003}{{\em Phys. Rev.} {\bf
  D89} (2014) no.~7, 073003},
\href{http://arxiv.org/abs/1401.4185}{{\tt arXiv:1401.4185 [hep-ph]}}.

\bibitem{Actis:2010gg}
{\bf Working Group on Radiative Corrections and Monte Carlo Generators for Low
  Energies} Collaboration, S.~Actis {\em et al.}, ``{Quest for precision in
  hadronic cross sections at low energy: Monte Carlo tools vs. experimental
  data},'' \href{http://dx.doi.org/10.1140/epjc/s10052-010-1251-4}{{\em Eur.
  Phys. J.} {\bf C66} (2010)  585--686},
\href{http://arxiv.org/abs/0912.0749}{{\tt arXiv:0912.0749 [hep-ph]}}.

\bibitem{Lees:2012cj}
{\bf BaBar} Collaboration, J.~P. Lees {\em et al.}, ``{Precise Measurement of
  the $e^+ e^- \to \pi^+\pi^- (\gamma)$ Cross Section with the Initial-State
  Radiation Method at BABAR},''
  \href{http://dx.doi.org/10.1103/PhysRevD.86.032013}{{\em Phys. Rev.} {\bf
  D86} (2012)  032013},
\href{http://arxiv.org/abs/1205.2228}{{\tt arXiv:1205.2228 [hep-ex]}}.

\bibitem{Branchina:2013jra}
V.~Branchina and E.~Messina, ``{Stability, Higgs Boson Mass and New Physics},''
  \href{http://dx.doi.org/10.1103/PhysRevLett.111.241801}{{\em Phys. Rev.
  Lett.} {\bf 111} (2013)  241801},
\href{http://arxiv.org/abs/1307.5193}{{\tt arXiv:1307.5193 [hep-ph]}}.

\bibitem{Andreassen:2016cvx}
A.~Andreassen, D.~Farhi, W.~Frost, and M.~D. Schwartz, ``{Precision decay rate
  calculations in quantum field theory},''
\href{http://arxiv.org/abs/1604.06090}{{\tt arXiv:1604.06090 [hep-th]}}.

\end{thebibliography}\endgroup

\end{document}